\documentclass{article}

 \usepackage[main, final]{neurips_2025}

\usepackage[utf8]{inputenc} % allow utf-8 input
\usepackage[T1]{fontenc}    % use 8-bit T1 fonts
\usepackage{hyperref}       % hyperlinks
\usepackage{url}            % simple URL typesetting
\usepackage{booktabs}       % professional-quality tables
\usepackage{amsfonts}       % blackboard math symbols
\usepackage{nicefrac}       % compact symbols for 1/2, etc.
\usepackage{microtype}      % microtypography
\usepackage{graphicx}
\usepackage{xcolor}         % colors
\usepackage{enumitem}
\title{Rethinking Recommendation Paradigms: From Pipelines to Agentic Recommender Systems}

\author{%
\textbf{Jinxin Hu}$^{1*}$,
\textbf{Hao Deng}$^{1*}$,
\textbf{Lingyu Mu}$^{2}$,\\
\textbf{Hao Zhang}$^{1}$,
\textbf{Shizhun Wang}$^{1}$,
\textbf{Yu Zhang}$^{1}$,
\textbf{Xiaoyi Zeng}$^{1}$\\[4pt]
$^{1}$Alibaba International Digital Commerce Group, Beijing, China\\
$^{2}$University of Chinese Academy, Beijing, China\\[4pt]
\texttt{jinxin.hjx@alibaba-inc.com},
\texttt{denghao.deng@alibaba-inc.com},
\texttt{mulingyu@iie.ac.cn},\\
\texttt{zh138764@alibaba-inc.com},
\texttt{shaoan.wsz@taobao.com},
\texttt{daoji@alibaba-inc.com},\\
\texttt{yuanhan@taobao.com}
\thanks{Equal contribution.}%
}

\begin{document}

\maketitle

\begin{abstract}
  Large-scale industrial recommenders usually follow a fixed multi-stage pipeline (recall, ranking, re-ranking) and have progressed from collaborative filtering to deep and large pre-trained models. However, both multi-stage and “One Model” designs are essentially static: models are black boxes, and system improvement depends on manual hypotheses and engineering, which is hard to scale under heterogeneous data and multi-objective business constraints. We propose an Agentic Recommender System (AgenticRS) that reorganizes key modules as agents. Modules are promoted to agents only when they form a functional closed loop, can be independently evaluated, and possess an evolvable decision space. For model agents, we sketch two self-evolution mechanisms: RL-style optimization in well-defined action spaces, and LLM-based generation and selection of new architectures and training schemes in open-ended design spaces. We further distinguish individual evolution of single agents from compositional evolution over how multiple agents are selected and connected, and use a layered Inner/Outer reward design to couple local optimization with global objectives. This provides a concise blueprint for turning static pipelines into self-evolving agentic recommender systems.
\end{abstract}

\section{Introduction}
\begin{figure}[htbp]

  \includegraphics[width=1.0\textwidth]{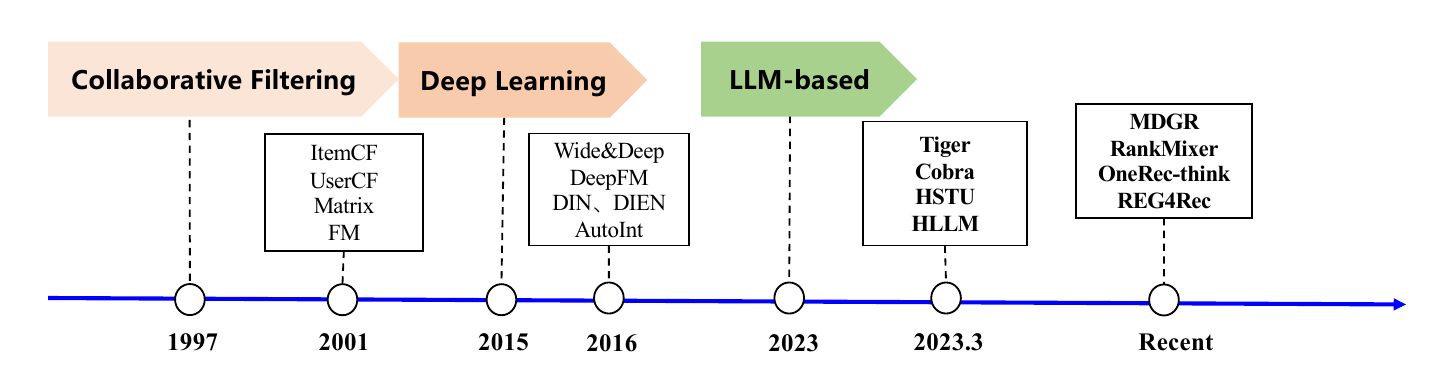}
  \vspace{-5pt}
  \caption{
  Technological Evolution of Recommendation Systems
  }
  % \Description{xx.}
  \label{fig1}
  \vspace{-0.5cm} %调整图片与上文的垂直距离
\end{figure}
\begin{figure}[htbp]

  \includegraphics[width=1.0\textwidth]{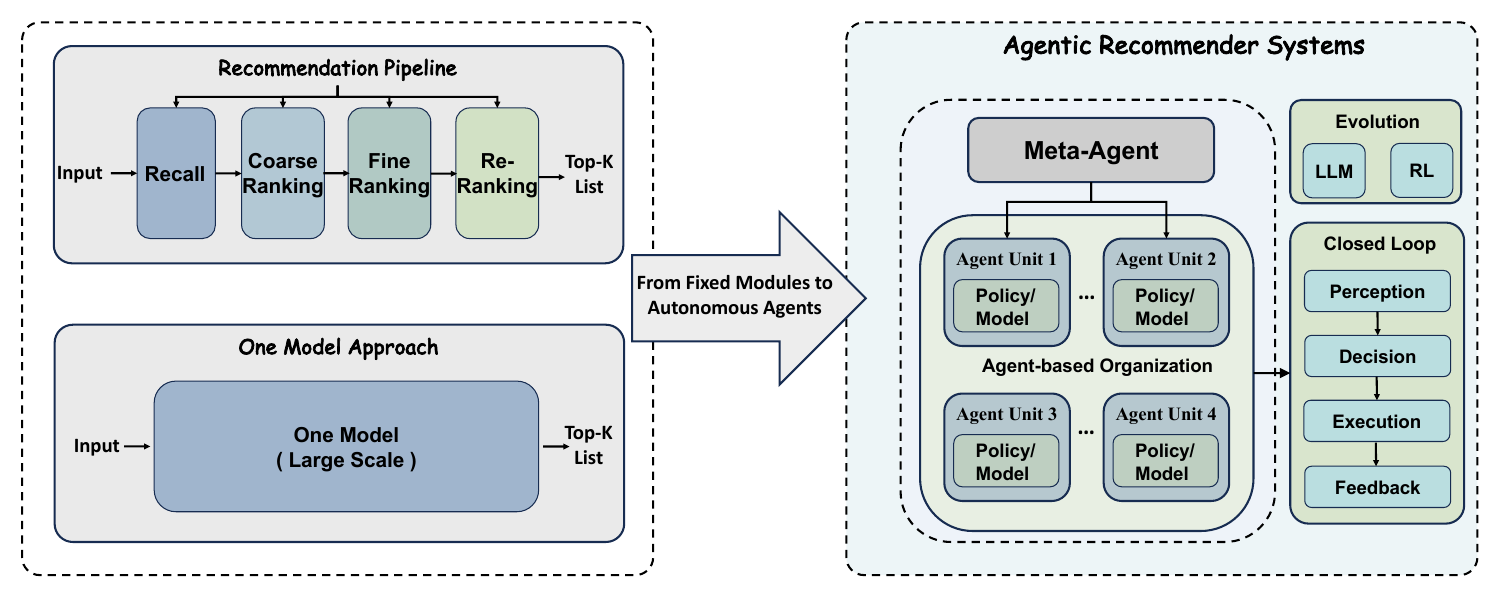}
  \vspace{-5pt}
  \caption{
  The Agentic Recommender Systems paradigm.
  }
  % \Description{xx.}
  \label{fig2}
  \vspace{-0.5cm} %调整图片与上文的垂直距离
\end{figure}

Large-scale recommender systems underpin search, feeds, short video, and e-commerce applications \cite{wang2021survey, wang2024rethinking, lin2024enhancing,wang2025home,mu2025trust,deng2025csmf}. Their models have progressed from neighborhood-based collaborative filtering and matrix factorization \cite{resnick1994grouplens,koren2009matrix} to deep neural networks \cite{cheng2016widedeep,guo2017deepfm,lian2018xdeepfm} and large pre-trained or generative models \cite{gray1984vector,rpg,zhou2025onerec,mu2025synergistic,wang2024learnable,lee2022autoregressive,wang2023generative,mu2026masked,xing2025reg4rec}, while system design has largely converged on multi-stage pipelines with multiple recall routes followed by coarse, fine, and re-ranking. However, as shown in Fig.~\ref{fig2}, these systems are still organized as static compositions of modules. Models act as black boxes, and progress depends on experts who manually adjust configurations, launch A/B tests, and interpret results, making rapid adaptation, localized capability improvement, and continuous autonomous evolution difficult. 

In this paper, we argue that recommender systems should be reframed as multi-agent decision systems rather than static model pipelines. Instead of viewing recall, ranking, and policy modules as fixed components, we propose to treat certain functionally closed, independently evaluable, and evolvable units as agents that operate in perception–decision–execution–feedback loops. Building on this agentic perspective, we outline an architecture-level view of Agentic Recommender Systems (AgenticRS), discuss concrete evolution mechanisms for such agents, and highlight how layered reward design can balance local optimization with global business alignment. Our goal is not to introduce a specific algorithm, but to provide a compact conceptual framework that can guide the transition from monolithic, manually tuned recommenders toward self-evolving, multi-agent recommendation systems.

\section{Why Agentic Recommender Systems?}

Reframing recommender systems as multi-agent decision processes is driven by concrete pressures that stress the static pipeline paradigm.

\textbf{Heterogeneous users, items, and scenarios.}
Modern platforms serve diverse users and rich content across many entry points. New and cold-start users coexist with heavy users; head items with long-tail and ephemeral content. A single model or a small fixed set of routes tends to favor dominant patterns while neglecting niche segments, motivating specialized components that can be activated and evolved differently across sub-distributions.

\textbf{Multi-objective, constraint-rich optimization.}
Industrial systems must balance short-term engagement, long-term value, ecosystem health, and risk or compliance constraints. Today these objectives are often entangled inside large models or implemented as ad hoc rules on top of scores, making it hard to attribute responsibility or adjust trade-offs in a principled way.

\textbf{Rising complexity and manual iteration cost.}
Production stacks already include many recall channels, ranking models, and policy components; adding large pre-trained or generative models further increases complexity. Yet evolution is still mostly manual: engineers diagnose failures, design changes, and run A/B tests. This process scales poorly and makes global effects of local tweaks hard to predict.

\textbf{Lack of continual, autonomous improvement.}
Although individual models can be retrained, the system as a whole lacks explicit mechanisms for self-improvement: it is unclear which units may adjust their structure or interactions, and there are no standard interfaces for evaluating and replacing them in isolation. The recommender thus behaves like a fixed production line rather than an adaptive entity.

These trends suggest that the core abstraction should shift from fixed modules in a pipeline to agents with explicit responsibilities and evolution capabilities. An agentic formulation enables assigning different agents to different subspaces and objectives, letting them learn within well-defined decision loops, and orchestrating their composition at the system level. The next section outlines an architecture-level view of such Agentic Recommender Systems.

\section{An Architecture-Level View of AgenticRS}
\label{sec:arch}

From an architectural perspective, an AgenticRS replaces a rigid stage-wise pipeline with a graph of interacting agents. Not every module is an agent; we only promote units that (i) participate in a functional closed loop, (ii) can be independently evaluated, and (iii) have an evolvable decision space. This section sketches how such agents are defined and organized.

\subsection{Functional and Model-Centric Agents}
We distinguish two broad classes of agents:
\paragraph{Functional agents.} These agents are defined by a closed-loop business function that may span multiple underlying models or tools. Examples include a traffic orchestration agent that segments users and routes requests across different pipelines, a strategy or experiment design agent that configures policies and allocates traffic, and a re-ranking policy agent that enforces diversity, freshness, or risk constraints. Their core responsibility is to decide how models are used, rather than to perform prediction themselves.

\paragraph{Model-centric agents.}
Model-centric agents encapsulate predictive or representation capabilities that can evolve in relative isolation. Examples include specific recall agents (e.g., behavior-based or content-based retrieval), ranking agents or specialized sub-towers for particular user/item segments, and fusion or calibration agents that combine scores from multiple models. These agents focus on improving local prediction quality under stable input--output interfaces and local evaluation metrics.

Each agent participates in a perception--decision- execution--feedback loop: it observes a well-defined state, produces an action (such as a candidate set, ranked list, or configuration), the system executes this action, and the resulting feedback is used to update the agent. Stable interfaces for states, actions, and feedback make it possible to replace or evolve agents without redesigning the entire system.

\subsection{Organizing Agents in the Recommendation Stack}
\label{sec:agent_layers}
\begin{figure}[htbp]

  \includegraphics[width=1.0\textwidth]{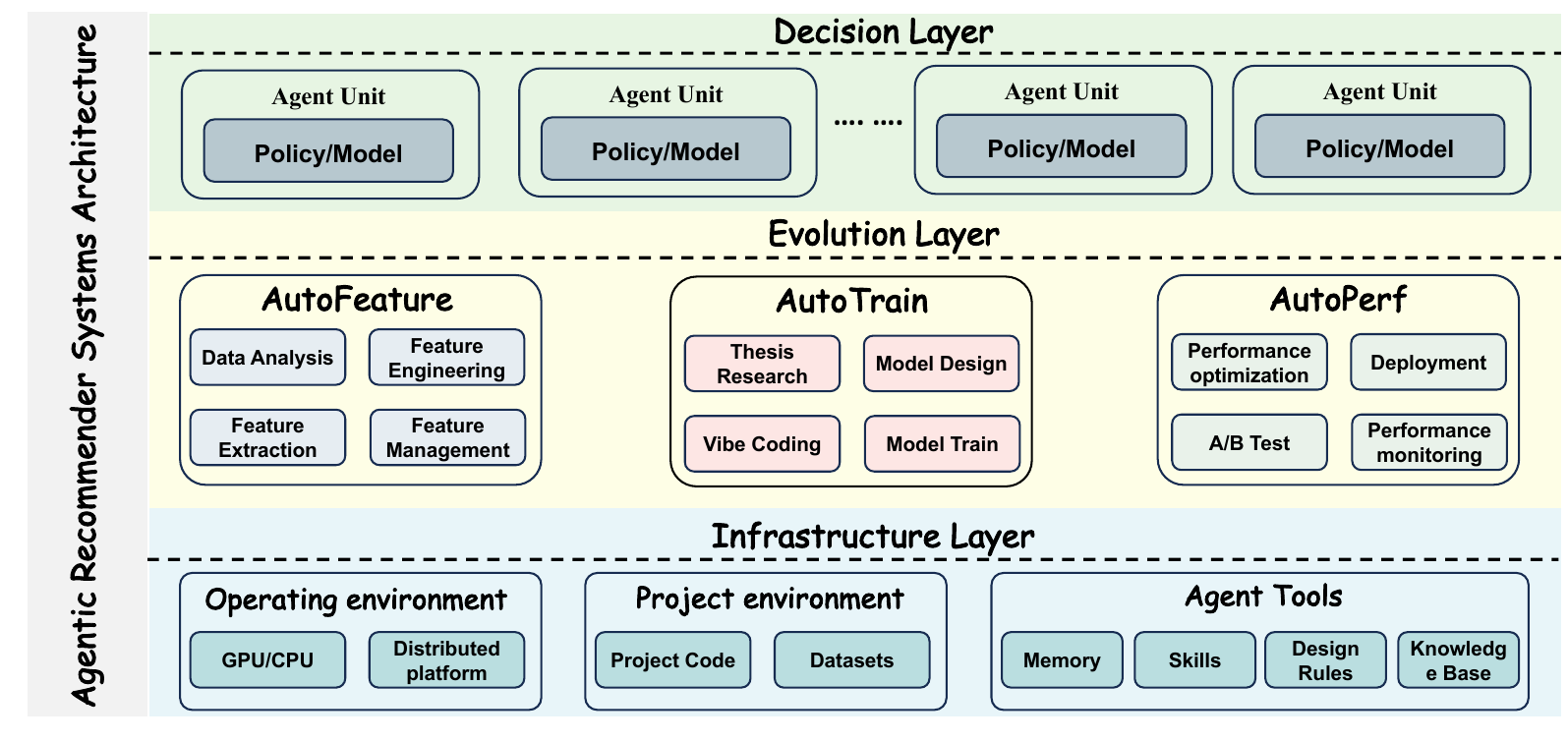}
  \vspace{-5pt}
  \caption{
  The architecure of agentic recommender systems.
  }
  % \Description{xx.}
  \label{fig_arch}
  \vspace{-0.5cm} %调整图片与上文的垂直距离
\end{figure}

At the system level, an AgenticRS can be organized into three cooperative layers of agents, each with a distinct role as shown in Figure \ref{fig_arch}.

\paragraph{Decision layer.}
The decision layer contains agents that directly make recommendation decisions for user requests. These agents inherit the responsibilities of traditional recall, ranking, re-ranking, and strategy-control modules, but their internal structures and combinations are no longer fixed. Given a request and shared state, decision agents produce actions such as candidate sets, ranked lists, or policy adjustments that determine the final exposure.

\paragraph{Evolution layer.}
The evolution layer hosts agents responsible for data analysis, model and policy design, training, and deployment. They consume logs and reward signals from the decision layer, and continuously propose and update new versions of decision agents. Outputs may include revised architectures, hyperparameters, routing rules, or strategy configurations, which are validated and rolled out via controlled experiments.

\paragraph{Infrastructure layer.}
The Infrastructure layer provides infrastructure for task orchestration and knowledge storage. Agents in this layer maintain unified state and experience for the other two layers, including user and item profiles, long-term interaction histories, global constraints, and meta-knowledge about past experiments. They support task scheduling across agents, resolve conflicts between competing objectives, and expose consistent interfaces for reading and writing system-wide memory.

This layered organization preserves the practical benefits of a multi-stage recommender, while turning key components into explicit agents that can be independently optimized, composed, and evolved. In the next section, we discuss the mechanisms that drive such evolution.

\section{Evolution Mechanisms for Agents}

Turning recommender modules into agents is only meaningful if they can improve over time. In AgenticRS, evolution refers both to how a single agent updates its architecture, hyperparameters, or decision policy from feedback, and to how the system updates the set and wiring of agents. We emphasize two mechanisms and two granularities.

\subsection{RL- and Search-Based Local Optimization}

For many model-centric agents, the design space can be described by a moderate number of discrete or continuous choices (backbone, loss weights, sampling ratios, learning rates, routing thresholds, etc.). In these cases, evolution can be cast as reinforcement learning or black-box search: an agent state summarizes the current data regime and constraints, an action specifies an architecture or hyperparameter configuration, and a reward comes from offline metrics or small-scale online tests. A controller learns to propose configurations with higher expected reward, reusing experience across iterations and reducing manual tuning cost under shifting distributions.

\subsection{LLM-Driven Structural Innovation}

Some agents operate in a much less parameterizable space, where they must redesign multi-tower structures, multi-task objectives, or combinations of recall and ranking paths. Here the action space is high-dimensional and structured, making it hard to enumerate options as RL actions. Large language models can act as design assistants: given descriptions of the current agent, logs of failures, historical experiments, and business constraints, an LLM proposes candidate architectures, training procedures, or routing strategies in natural language and code. These candidates are instantiated and evaluated with standard pipelines; successful variants are retained, and their designs are stored in the shared knowledge base as experience for future iterations.

\subsection{Individual vs.\ Compositional Evolution}

Agent evolution happens at two levels. \textbf{Individual evolution} assumes a fixed system graph and improves a single agent while others stay roughly unchanged, using RL or search for fine-grained configuration tuning and LLMs for larger architectural edits. \textbf{Compositional evolution} changes which agents exist and how they are connected, for example selecting recall agents, reconfiguring ranking ensembles, or adjusting routing for different user segments. It can also be driven by search, RL, or LLM proposals, evaluated against global objectives such as business metrics and resource budgets. In practice these two levels alternate: given a composition, individual agents are optimized; when gains saturate or conditions shift, the system explores new compositions and then resumes local refinement under the updated architecture.

\section{Conclusion}
This paper proposes a conceptual framework and design principles for Agentic Recommender Systems (AgenticRS), aiming to move from static, manually tuned architectures to self-evolving ecosystems. We define model-centric agents using three criteria: functional closure, independent evaluability, and evolutionary potential, providing a structured way to decentralize model intelligence. Agent evolution is organized along two axes: methodological, combining RL based and LLM based mechanisms for autonomous optimization; and structural, distinguishing individual evolution of single agents from the compositional evolution of multi agent networks. A hierarchical reward design underpins this process: most sub agents primarily optimize Inner Rewards for stable local competence, while key coordination agents are driven by Outer Rewards to align the overall system with business objectives. Rather than building monolithic models, this perspective emphasizes designing adaptive environments in which agents interact and improve, offering a compact blueprint for sustainable industrial scale recommender systems.

\bibliographystyle{plain}
\bibliography{references}
\end{document}